# SMALLEST COVERING REGIONS AND HIGHEST DENSITY REGIONS FOR DISCRETE DISTRIBUTIONS

BEN O'NEILL[*], *Australian National University*[**]

WRITTEN 1 SEPTEMBER 2020; REVISED 25 SEPTEMBER 2021

**Abstract**

This paper examines the problem of computing a canonical smallest covering region for an arbitrary discrete probability distribution. This optimisation problem is similar to the classical 0-1 knapsack problem, but it involves optimisation over a set that may be countably infinite, raising a computational challenge that makes the problem non-trivial. To solve the problem we present theorems giving useful conditions for an optimising region and we develop an iterative one-at-a-time computational method to compute a canonical smallest covering region. We show how this can be programmed in pseudo-code and we examine the performance of our method. We compare this algorithm with other algorithms available in statistical computation packages to compute HDRs. We find that our method is the only one that accurately computes HDRs for arbitrary discrete distributions.

COMPUTATION; DISCRETE DISTRIBUTION; HIGHEST DENSITY CONDITION; SMALLEST COVERING REGION; 0-1 KNAPSACK PROBLEM.

## 1. Introduction

Statistical prediction problems sometimes require the analyst to find the smallest region on the support of a random variable that has some specified minimal coverage probability. Smallness is measured with respect to the dominating measure for the probability measure of the random variable. Usually this problem is considered for continuous distributions that are dominated by Lebesgue measure, and in this context, the smallest covering region is (almost) equivalent to the "highest density region" (HDR), which is obtained by taking the set of all values with density at least as large as some cut-off value (Box and Tiao 1973, Hyndman 1996). By taking this set of "highest density" points, one obtains a set that is a smallest covering region with minimum coverage probability equal to its actual coverage probability. Non-equivalence of the concepts arises from the fact that the possible coverage probabilities for the HDR can have "jumps" as we raise the density cut-off, owing to flat sections in the density.[1] If the minimum required coverage probability is inside a gap from this "jump" then a smaller covering region can be obtained by removing a section of the values in the HDR which have the smallest density above the cut-off value.

---

In any case, statistical literature on this problem focuses on HDRs for continuous distributions, where the task of finding the boundaries of the region is solved using calculus techniques or grid methods, depending on the dimensionality of the problem and the shape of the distribution (Hyndman 1996). In the present paper we examine the underappreciated problem of computing the smallest covering region for an arbitrary discrete probability distribution (i.e., a distribution with countable support). At first glance this may seem like a trivial problem; sorting methods can order probability values, and so we can determine the values with the highest probability within a set, and take these in descending order of probability until we have a smallest covering region. However, from a computational perspective the problem is rendered non-trivial by the fact that we can only sort probability values over finite sets, and many discrete distributions have countably infinite support. This complication renders the computational problem non-trivial and behoves an investigation of appropriate methods for the problem.

In practice, most families of discrete probability distributions used in statistical problems are univariate and unimodal (quasi-concave) distributions over the integers, and in this case, there will be a smallest covering region that is a contiguous set of integers. In such cases, finding the smallest covering region reduces to finding the two boundary points of an interval, and this can be done efficiently using standard discrete calculus methods. However, in this paper we will examine the general problem of computing the smallest covering region for an arbitrary discrete distribution with no structural assumptions beyond the fact that the support is known to be a subset of some known countable set of values (for brevity, we say that the distribution is "concentrated on" a known countable set). We will show how the concept of a smallest covering region relates to the concept of a highest density region for discrete distributions, and we will formulate methods for finding both.

Highest density regions are of interest in Bayesian analysis in the formulation of "credibility regions" and "highest posterior density regions" (see e.g., Wei and Tanner 1990; Turkkan and Pham-Gia 1993; Chen and Shao 1999) and also for broader classical inference problems using "highest confidence density regions" (see e.g., Tian *et al* 2011). These regions are also used in forecasting problems to obtain a "prediction region" for an observable variable (see e.g., Hyndman 1995; Kim, Fraser and Hyndman 2011). In many statistical problems, interest will focus on a continuous observable value or unknown parameter, but there are certainly problems where a discrete random variable or parameter may be of interest, and in such cases the discrete version of the HDR is of interest.

In addition to being a useful problem in computational statistics in its own right, this problem is also interesting, since it yields an optimisation problem that is a kind of "inverse" variation of the classical "knapsack problem" in computer science. This latter problem is a discrete optimisation where we fill a knapsack of a fixed weight capacity with items having different weights and values. The problem of finding a discrete HDR turns out to be an inverse variant of this problem with a countably infinite set of objects. This is an interesting variant of a well-known optimisation problem, requiring us to develop a method to deal with a countably infinite set of objects with a finite algorithm. We will present an algorithm for efficient computation of the solution to this optimisation problem, using a technique that is useful both in the present case and also in broader optimisation problems taken over countably infinite spaces.

The two purposes of this paper are to establish some useful mathematical theory for HDRs for discrete distributions, and to create an effective algorithm to compute the HDR for an arbitrary discrete distribution, even if the support is countably infinite. From a practical perspective, this manifests in an automated algorithm for computation of HDRs in discrete distributions that can use used by statistical practitioners without concern for the underlying optimisation problem. Our analysis is implemented in user-friendly form in the `stat.extend` package in `R` (O'Neill and Fultz 2020), allowing statistical practitioners to easily compute the HDR for any discrete distribution. To the knowledge of the present author, this function is presently the only available facility that accurately computes the HDR in all cases.

**2. Smallest covering regions for discrete distributions**

We will consider an arbitrary **discrete probability measure** $\mathbb{P}$ on a countable sample space $\Omega$. As is standard in probabilistic analysis, we will assume that this sample space is known, so that the support of the distribution is at least narrowed down to a known countable set.[2] To facilitate our analysis, we denote the probability mass function for our measure by $f$ (formally this is the Radon-Nikodym derivative of $\mathbb{P}$ with respect to counting measure), which allows us to write the probability of any event as:

$$\mathbb{P}(\mathcal{A}) = \sum_{x \in \mathcal{A}} f(x) = \sum_{x \in \Omega} f(x) \mathbb{I}(x \in \mathcal{A}) \qquad \text{for } \mathcal{A} \subseteq \Omega.$$

---

[2] In the unusual event that the support is not concentrated on a known countable set, the main challenge becomes the detection of the discrete support within an uncountable space. That problem is beyond the scope of this paper and is unlikely to arise in any realistic probabilistic or statistical analysis.

We note that we can also write the size of any event (with respect to counting measure) as:

$$|\mathcal{A}| = \sum_{x \in \mathcal{A}} 1 = \sum_{x \in \Omega} \mathbb{I}(x \in \mathcal{A}) \qquad \text{for } \mathcal{A} \subseteq \Omega.$$

We further define the following functions giving the smallest and largest mass over a set:

$$\nabla(\mathcal{A}) \equiv \inf_{x \in \mathcal{A}} f(x) \qquad \Delta(\mathcal{A}) \equiv \sup_{x \in \mathcal{A}} f(x).$$

For any $0 \leq \alpha \leq 1$ a **smallest covering region (SCR)** with minimum coverage probability $1 - \alpha$ is any solution to the following optimisation problem:

$$\text{Minimise } |\mathcal{A}| \qquad \text{subject to } \mathbb{P}(\mathcal{A}) \geq 1 - \alpha.$$

The optimisation problem for the smallest covering region always has at least one solution, but there are cases where multiple solutions can occur. For a discrete distribution, this occurs when there are multiple regions containing the same number of elements that all meet the minimum coverage probability requirement in the optimisation, but removal of any element from any of those regions would reduce its probability below the required coverage probability. This is illustrated in Figure 1 for the binomial distribution with size $n = 10$ and probability $\theta = 0.52$.

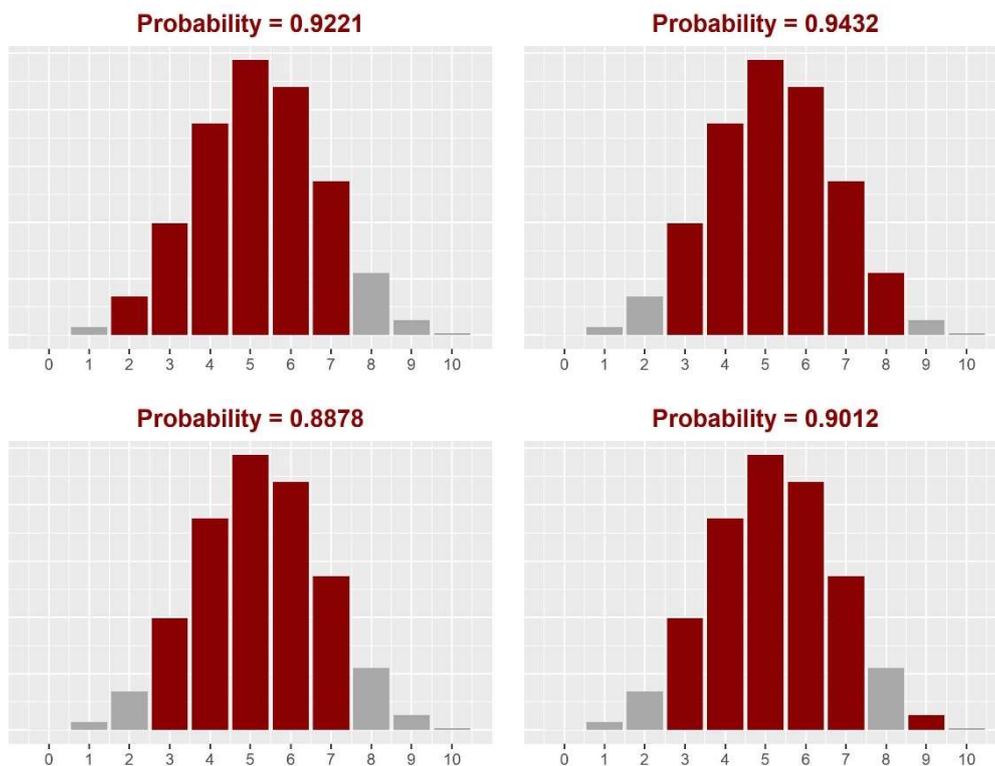

**FIGURE 1:** Covering regions for the $\text{Bin}(10, 0.52)$ distribution

(All but the bottom-left plot are solutions to the 90% smallest coverage region problem;
observe that the top-right plot is the "canonical" solution with the highest coverage probability)

One interesting aspect of the plots in Figure 1 is that, of the three solutions giving a smallest covering region with 90% minimum coverage probability, the top-right solution is the one with the highest coverage probability. It can therefore be regarded as the "canonical" solution that gives the best trade-off between size and coverage while still meeting the minimum coverage requirement. For statistical prediction purposes, this is the solution of most interest. Indeed, if we look at the goal of statistical prediction, it should be clear that this particular solution is the best prediction region, insofar as it gives the highest coverage probability among all the regions of the same size. It not only solves the optimisation problem for the smallest covering region; it also solves the "dual problem" of maximising the coverage probability among all regions of that "optimal" size.

In view of this, it is desirable to reframe our analysis of smallest covering regions to look at the dual problem where we seek to maximise the coverage probability among all regions with the "optimal" size (i.e., the smallest size needed to meet the minimum coverage probability), yielding a "canonical" solution to the problem. To do this, let us define the objects:

$$R(\alpha) \equiv \min_{\mathcal{A} \in \mathfrak{C}(\alpha)} |\mathcal{A}| \qquad \mathfrak{C}(\alpha) \equiv \{\mathcal{A} \subseteq \Omega | \mathbb{P}(\mathcal{A}) \geq 1 - \alpha\}.$$

The class $\mathfrak{C}(\alpha)$ contains all the regions that give minimum coverage probability $1 - \alpha$, and the value $R(\alpha)$ is the size of each smallest covering region. Our original optimisation problem was to find sets in $\mathfrak{C}(\alpha)$ with minimum size. We can now state the dual problem leading to the canonical solutions. For any $0 \leq \alpha \leq 1$ a **canonical smallest covering region (CSCR)** with minimum coverage probability $1 - \alpha$ is any solution to the following optimisation problem:

$$\text{Maximise } \mathbb{P}(\mathcal{A}) \qquad \text{subject to } |\mathcal{A}| \leq R(\alpha).$$

This dual problem is the optimisation that determines the "canonical" solutions for the smallest covering region. Every solution to this dual problem is also a solution to the primal problem stated above, but not every solution of the latter is a solution of the former. In particular, since this dual problem maximises the probability among all sets with $|\mathcal{A}| \leq R(\alpha)$, it will exclude all solutions to the primal problem where the coverage probability is less than in another solution. In the example shown in Figure 1, the region shown in the top-right plot is the only solution to this dual problem, and so it is a "canonical" solution to the smallest covering region with the stipulated minimum coverage probability.

This optimisation problem is an "inverse" variation of the classical "0-1 knapsack problem", which is a famous combinatorial optimisation that has spawned a wide range of algorithms in computer science (see e.g., Marello and Toth 1990, Kellerer, Pferschy and Pisinger 2004). In that problem we have a finite set of "items" $x = 1, \ldots, n$ with weights $w_x$ and values $v_x$ and we have a "knapsack" with weight capacity $W$. We wish to find the set of items with the highest total value subject to the requirement that they fit in the knapsack (i.e., their total weight must not exceed the weight capacity). In the classical problem there are a finite number of items and all these items have weights and values that are positive integers. However, we can vary the problem to allow a countably infinite set of items $x \in \Omega$ with real weights $w_x \in \mathbb{R}$ and real values $u_x \in \mathbb{R}$. The resulting optimisation problem is:

$$\text{Maximise} \sum_{x \in \Omega} u_x \mathbb{I}(x \in \mathcal{A}) \qquad \text{subject to} \sum_{x \in \Omega} w_x \mathbb{I}(x \in \mathcal{A}) \leq W.$$

With some simple algebra we can easily confirm that the optimisation problem for a canonical smallest covering region is an instance of this optimisation problem using a "knapsack" with weight capacity $W = R(\alpha)$, item weights $w_x = 1$ and item values $u_x = f(x)$. This problem is the "unit weight" version of the 0-1 knapsack problem, but with two differences. Firstly, we have a countable number of items (which may be infinite) rather than a finite number of items. Secondly, the weight capacity of the knapsack is not a value that is known in advance — it is itself determined as the solution to the primal optimisation problem, which means that this variant of the problem is a kind of "inverse" knapsack problem.

As with the primal optimisation problem, our dual optimisation problem can lead to multiple canonical solutions. This occurs when there are several equiprobable values in the distribution and we need to include some —but not all— of those values in the region giving a canonical solution. This is illustrated in Figure 2 for the binomial distribution with size $n = 10$ and probability $\theta = 0.50$. In this example, the two canonical solutions occur by swapping out one of two equiprobable values that occur as the lowest probability values in the region, only one of which is required to meet the minimum coverage probability. Although we defer formal proof until later, we can see from Figures 1-2 that the canonical smallest covering regions are all "highest density regions", in the sense that elements outside the region have probability no greater than elements in the region. Note that this is a "weak" highest density condition, insofar as it allows values outside the region to be equiprobable to values in the region. It is therefore a different conception of a "highest density region" than is obtained by taking the set of all values that meet some minimum density cut-off.

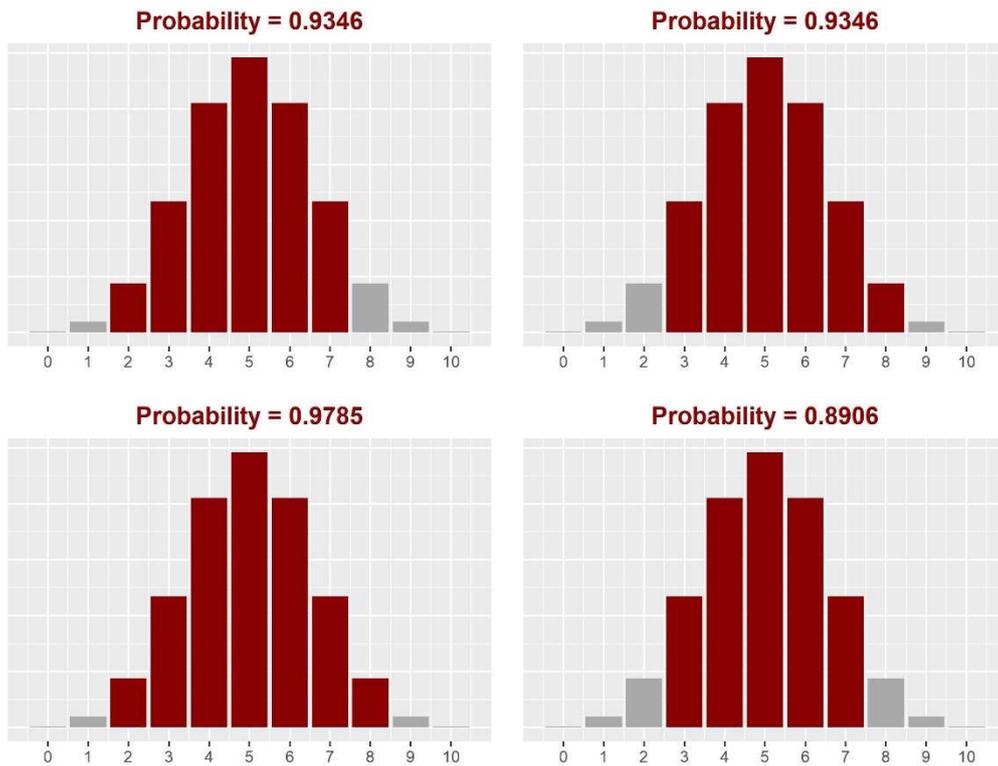

**FIGURE 2:** Covering regions for the $\text{Bin}(10, 0.50)$ distribution

(The plots at the top show two canonical solutions to the 90% smallest coverage region problem;
the plots at the bottom show two regions with inadequate/excess coverage probability)

In order to identify all the canonical smallest covering regions (i.e., all solutions to the dual optimisation problem set out above), it is sufficient to compute a single solution, and identify the set of all values in the support that are equiprobable with the least probable value in the canonical smallest covering region (which we will call the "variation set"). This allows us to form other solutions by "swapping out" elements in the variation set in such a way that a fixed number of elements are included. In the example shown in Figure 2, it would be sufficient to find the solution shown in the top-left plot (which includes $x = 2$ but excludes $x = 8$), but also identify the variation set composed of the elements $x = 2$ and $x = 8$. The latter solution can then be obtained by swapping the points in the variation set to yield the region shown in the top-right of the figure. In general, if there are $n$ values in the variation set and $r$ of these values must be included to obtain the minimum coverage probability of the region then there will be $\binom{n}{r}$ solutions to the optimisation problem. All of these solutions have the same coverage probability and the same size, so none is statistically more favourable than another. Indeed, statistical prediction problems usually require only a single solution to the problem, and so it is usual that we would not need to compute all solutions.

# 3. One-at-a-time iterative method for computing CSCR/HDR

We have already noted that the optimisation problem for the discrete CSCR/HDR is a variant of the classical "0-1 knapsack problem", which is a weak NP-hard problem. This optimisation problem has several variants in the literature and a number of exact and approximate solution methods (for an overview, see Martello and Toth 1990; Pisinger 2005). As a prelude to our discussion of the present optimisation of interest, we begin with a brief excursion into the general mathematical methods that have been used to solve that problem. There are a range of well-known algorithms for computing the optimising set in the classical problem.

Markov Chain Monte Carlo (MCMC) methods can be used to give approximate solutions for the knapsack problem. One approach to the optimisation is the use of "simulated annealing" which is a stochastic optimisation method similar to the Metropolis-Hastings MCMC method. The simulated annealing method explores the surface of the objective function stochastically using the Metropolis acceptance-rejection method. The algorithm terminates when the average change in the objective function is small, when it rejects a large number of consecutive points, or at other stopping conditions such as a maximum number of total iterations or running time. The simulated annealing method can be "parallelised" by producing parallel runs of candidate points and exchanging solution information either at all or occasional iterations. Regular and parallel simulated annealing in the 0-1 knapsack problem have been examined in a number of papers, and has been subjected to comparative analysis with other methods (see e.g., Lin, Kao and Hsu 1993; Sonuc, Sen and Bayir 2016; de Almeida Dantas and Cáceres 2018; Moradi, Kayvanfar and Refiee 2021).

Another class of methods are "genetic" or "evolutionary" algorithms that solve optimisation problems using a process that mimics evolutionary "natural selection" processes in biology (Goldberg 1998). These algorithms give a stochastic process converging to the solution of an optimisation problem by simulating an evolution where each iteration in the search space is treated as a "generation" that evolves using either a fitness function or some other evolutionary specification. The broad nature of this method leads to a variety of specific algorithms. Genetic algorithms have been applied to a range of knapsack problems (see e.g., Khuri, Bäck and Heitkötter 1994; Chu and Beasley 1998; Gupta and Garg 2009; Changdar, Mahapatra and Pal 2015; Rezoug, Bader El-Den and Boughaci 2018; Ali, Essam and Kasmarik 2021).

The stochastic nature of these algorithms means they are not guaranteed to produce an exact solution, but the discrete search space for the 0-1 knapsack problem has a non-zero probability of an exact solution.[3] Dynamic programming methods can be used to solve the 0-1 knapsack problem, by breaking the optimisation problem down into recursive steps involving simpler problems (usually with heavier constraints). Dynamic programming has been used to solve the 0-1 knapsack problem (see e.g., Martello, Pisinger and Toth 1999). Other exact algorithms are available for the classical version of the problem (Martello, Pisinger and Toth 2000). Unfortunately, existing methods of this kind are only available when the weight capacity is known in advance, and the set of items is finite, which is not the case in our problem.

Before presenting our method to compute a solution for the canonical smallest covering region, we will show two useful characterisations of this region. Theorem 1 shows that a canonical smallest covering region can be characterised by the minimum coverage requirement plus "moderation" and "highest density" conditions. It is notable that the highest density condition is a weaker condition than taking all points greater than or equal to some given density cut-off. Instead, it requires only that the smallest density of values in the region is at least as large as the largest density of values outside the region. This allows equiprobable values to occur on the margin, with some of those values included in the canonical smallest covering region and others excluded from the region. The moderation condition requires the removal of any points from the region that would put its total probability below the minimum coverage probability.

**THEOREM 1:** Suppose that $\mathbb{P}$ is a discrete probability measure concentrated on a countable set $\Omega$. Then, subject to an extremely minor exception set out in the footnote,[4] a set $\mathcal{H}$ is a canonical smallest covering region (CSCR) with minimum coverage probability $1 - \alpha$ if and only if the following conditions are satisfied:

$$
\begin{aligned}
&\text{(Minimum coverage)} &&\mathbb{P}(\mathcal{H}) \geq 1 - \alpha, \\
&\text{(Moderation)} &&\mathbb{P}(\mathcal{J}) < 1 - \alpha \ \text{ for all } \mathcal{J} \subset \mathcal{H}, \\
&\text{(Highest density)} &&\nabla(\mathcal{H}) \geq \Delta(\Omega - \mathcal{H}).
\end{aligned}
$$

---

[3] The reader may naturally ask, what is the probability of an exact solution from the algorithm? Unfortunately, computing this probability is even harder than computing the solution to the knapsack problem, so it is not useful to enquire into this probability in the present context.

[4] The exception occurs in the case where $\alpha = 0$ (i.e., a 100% covering region) where the support of the distribution is a canonical smallest covering region. If the support of the distribution is countably infinite then any countable superset of the support is also "of the same size" and is therefore also a canonical smallest covering region, even if it is a strict superset that does not obey the moderation condition.

Theorem 2 gives an alternative characterisation of the highest density condition that is useful for computational purposes. The theorem shows that this condition is satisfied if we can find a "search set" that satisfies the inner and outer boundedness conditions. In practice we need the search set to be finite, so that it is possible to confirm the two boundedness conditions with finite number of computations. In cases where the sample space $\Omega$ is (countably) infinite we check the highest density condition by using a finite search set $\mathcal{S}$ and checking inner and outer boundedness to establish the highest density condition. This is the basis for our algorithm later in the paper.

**THEOREM 2:** Suppose that $\mathbb{P}$ is a discrete probability measure concentrated on a countable set $\Omega$. Then a set $\mathcal{H}$ satisfies the highest density condition in Theorem 1 if and only if there exists a set $\mathcal{H} \subseteq \mathcal{S} \subseteq \Omega$ (called the "search set") such that the following conditions are satisfied:

$$\text{(Inner-boundedness)} \quad \nabla(\mathcal{H}) \geq \Delta(\mathcal{S} - \mathcal{H}),$$
$$\text{(Outer-boundedness)} \quad 1 - \mathbb{P}(\mathcal{S}) \leq \nabla(\mathcal{H}).$$

In the case where $\nabla(\mathcal{H}) > 0$ there exists a finite search set $\mathcal{S}$ satisfying these conditions.

We are now in a position to present an iterative computational method for finding a canonical smallest covering region. We consider the elements in a given sequence $\Omega = \{\omega_1, \omega_2, \omega_3, \ldots\}$ and iteratively compute a "candidate region" over each search set $\mathcal{S}_k = \{\omega_1, \ldots, \omega_k\}$, with the value $k$ increasing by one unit each time we iterate. At each iteration the candidate region is computed in a way that ensures that it meets the coverage, moderation, and inner-boundedness conditions for the current search set. The iterations continue over the sequence until the outer-boundedness condition is satisfied, and at this point the algorithm terminates and the candidate region is a canonical smallest covering region. For any value $\alpha > 0$ (i.e., for any coverage probability less than one), the algorithm will terminate in a finite number of iterations. The number of iterations required depends on the sequence used — since the algorithm terminates upon satisfaction of the outer-boundedness condition, it will tend to run faster if this sequence runs through high-probability values earlier than low-probability values.

Since this algorithm iteratively forms a canonical smallest covering region from the first $k$ elements, the first iterations add elements to the candidate region without displacing existing elements. This occurs up until the point where the candidate region has a total probability that is above the minimum coverage probability. After this point, addition of any new point to the candidate region must displace at least one existing element. Since the initial iterations add elements without displacing any existing elements, it is efficient to start the algorithm by first computing the smallest value $K$ such that $\sum_{i=1}^{K} f(\omega_i) \geq 1 - \alpha$. We begin the algorithm with the baseline search set $S_K \equiv \{\omega_1, \ldots, \omega_k\}$ and with all elements included in the candidate region; later iterations then proceed from this point by considering one new element at a time. Before proceeding to the pseudo-code for our algorithm, we briefly describe the representation and updating of the "candidate region" at each iteration. For ordering purposes, we have found it best to represent this object as a vector composed of complex values $\omega_i + \sqrt{-1} \cdot f(\omega_i) \in \mathbb{C}$. For a candidate region with $r$ elements, we use a complex vector **CANDIDATE** of length $r$, with real parts containing the elements in the region and imaginary parts containing their probabilities. Elements are arranged in descending probability order (i.e., descending order on the imaginary part), so the lowest-probability elements occur at the end of the complex vector. The vector begins with the first $K$ elements included (which is also the search set), which means that the coverage, moderation and inner-boundedness conditions all hold.

Our algorithm takes an input `cover.prob` giving the minimum coverage probability for the smallest covering region. At each iteration we update the candidate region in **H** and also keep track of the smallest probability of all elements in this region (**MINPROB**), the probability of the new element in the search set (**NEWPROB**), and the total probability of all elements that are not in the search set (**OUTPROB**). Each iteration adds one new element to the search set and we update the candidate region and other values by the following process.

- **Step 1:** Compute **NEWPROB** by applying the density function to the new element.
- **Step 2:** Update the value **OUTPROB** (by subtracting **NEWPROB** from its existing value)
- **Step 3:** If **NEWPROB > MINPROB** then the new element has a higher probability than the lowest-probability element in the candidate region. In this case the new element is added to **CANDIDATE** in the appropriate position, displacing low-probability elements from the end of this vector until the moderation requirement is met. This ensures that the coverage, moderation, and inner-boundedness requirements remain satisfied. We then update the value **MINPROB** from the updated **CANDIDATE** vector.
- **Step 4:** Check the outer-boundedness condition (**OUTPROB <= MINPROB**). If this condition is satisfied then the algorithm terminates; otherwise we start a new iteration that considers a new element in the search set.

Algorithm 1 below shows our method in pseudo-code. This algorithm takes a probability mass function `f` and the minimum coverage probability `cover.prob` and computes a canonical smallest covering region, presented in matrix form. The probability mass function is assumed to be defined over an arbitrary countable set Ω, and we let `E` be a function defining a sequence over this set (i.e., a bijective function mapping the positive integers to the set Ω). (We allow the function `E` to be given as an optional input to our algorithm, but with a default if this is not specified.) Steps 1-3 above occur in the `while` loop and the loop condition constitutes the check in Step 4. The output is the vector of values in the CSCR.

```
              ALGORITHM 1: Canonical Smallest Covering Region
        (arbitrary discrete distribution over a countable sequence Ω)
```

```
Input:      Probability density function f defined over x ϵ Ω
            Minimum required coverage probability cover.prob
            Bijective function E (optional — default to NULL)
Output:     The canonical smallest covering region for the density f
            (given as a vector containing the elements of this region)
```

```
#Set a sequence function (if not specified by the user)
if (is.null(E)) { E <- function(x) { Output element x in sequence Ω } }

#Compute initial candidate matrix and values
K         <- Smallest value with f(E(1)) + … + f(E(K)) ≥ cover.prob
CANDIDATE <- Complex vector with length K with elements
                CANDIDATE[i] <- E(i) + (Imaginary Unit)*f(E(i))
CANDIDATE <- Sort CANDIDATE into descending order on its imaginary part
MINPROB   <- Im(CANDIDATE[K])
OUTPROB   <- 1 – sum(Im(CANDIDATE))

#Iteratively update the candidate matrix
ITER <- 0
while (MINPROB < OUTPROB) {
  ITER    <- ITER + 1
  x       <- K + ITER
  NEWPROB <- f(E(x))
  OUTPROB <- OUTPROB - NEWPROB
  if (NEWPROB > MINPROB)  {
    #Add new element to candidate region
    T        <- Smallest value with NEWPROB > Im(CANDIDATE[T])
    NEWELEM  <- E(x)+ Imaginary unit *NEWPROB
    NEWCAND  <- Copy of CANDIDATE with NEWELEM inserted as Tth element
    #Prune the candidate region
    R        <- Smallest value with sum(Im(NEWCAND[1:R])) ≥ cover.prob
    CANDIDATE <- NEWCAND[1:R]
    MINPROB  <- Im(CANDIDATE[R]) } }

#Give output – vector of values for the CSCR (sorted)
sort(Re(CANDIDATE))
```

The above algorithm will successfully compute a canonical smallest covering region for any discrete distribution defined over the input set Ω. So long as the coverage probability is less than one, the algorithm will always terminate in a finite number of steps.[5] This gives us an "in principle" guarantee of feasibility of the algorithm, subject to computational power/memory. The efficiency of the algorithm depends primarily on how well the sequence function **E** travels through the space of the distribution — so long as this sequence function covers the highest probability elements of the distribution early, and then goes into the tails later, the algorithm will tend to compute the smallest covering region within a feasible number of operations. In practice, the discrete distributions in common use are concentrated on the integers, and so the sequence function **E** will be a bijection from the set of positive integers ℕ to the set of all integers ℤ. Moreover, if the support of the distribution is bounded, it is common that the user will know the bounds, and can incorporate this information into the chosen sequence function.

In Algorithm 2 we give a function that constructs a simple sequence function **E** for a discrete distribution on the integers, with an allowance for the user to specify bounds on the support. If the support is bounded from below then the sequence function starts at the lower bound and travels upward. If the support is bounded from above but not below then the sequence function starts at the upper bound and travels downward. If the support is unbounded then the sequence function starts at zero and travels outward in both directions in an oscillating fashion. This sequence function tends to perform reasonably well for most discrete distributions — the main exceptions occur for discrete distributions with unbounded support where the main "mass" of the distribution is far from zero.

In practice, Algorithm 2 can be subsumed into the first part of Algorithm 1 (where the sequence function is set) so that there is a default behaviour set for the sequence function in the absence of a user input. In this case Algorithm 1 takes the optional inputs `supp.min` and `supp.min` and these are used to determine the default behaviour of **E**. For most discrete distributions in statistical programming languages the user will have an available quantile function (in addition to the density function) and this can often be used to find bounds on the support. For example, in `R` we set `supp.min = Q(0)` and `supp.max = Q(1)` where `Q` is the quantile function for the distribution (and this also works fine when the bounds are infinite). This restricts the search to an interval that contains the support and which is bounded in some cases, making our search faster.

---

[5] For the special case where the coverage probability is one, the smallest covering region is the support of the distribution, and there is no algorithm that can compute this in a finite number of steps for an arbitrary distribution. For this reason, we recommend that the algorithm should implement this as a special case. We recommend that the function should return the entire set Ω in this case, and accompany this output with a warning message alerting the user that the returned region might not be the smallest covering region.

```
        ALGORITHM 2: A simple sequence function over the integers

Input:      Support lower bound supp.min (optional --- default to -Inf)
            Support upper bound supp.max (optional --- default to +Inf)
Output:     Bijective function E mapping positive integers to integers
```
```r
#Give supp.min and supp.max as integers
supp.min <- ceiling(supp.min)
supp.max <- floor(supp.max)

#Finite support (take values from left to right)
if ((supp.min >  -Inf)&(supp.max <  Inf)) {
  E <- function(i) { supp.min - 1 + i; } }

#Left-bounded support (take values from left to right)
if ((supp.min >  -Inf)&(supp.max == Inf)) {
  E <- function(i) { supp.min - 1 + i; } }

#Right-bounded support (take values from right to left)
if ((supp.min == -Inf)&(supp.max <  Inf)) {
  E <- function(i) { supp.max + 1 - i; } }

#Unbounded support (take values oscillating out from zero)
if ((supp.min == -Inf)&(supp.max == Inf)) {
  E <- function(i) { ifelse(i%%2 == 1, floor(i/2), -i/2); } }

#Give output
E
```

The above algorithms give a basic method for computation of the smallest covering region. It is possible to tweak the algorithms to add some additional user-friendly features if desired. For example, once a canonical smallest covering region has been computed, it is simple to add an additional step that searches over the finite search set to find all the elements in the validation set (i.e., all elements with `NEWPROB = MINPROB`), and this latter set can be included as part of the output to allow all canonical smallest covering regions to be obtained. It is also possible to program the algorithm to go one step further than this, and output the list of all the canonical smallest covering regions by substituting elements in and out of the validation set over all $\binom{n}{r}$ combinations. This is unlikely to be necessary in practice, since statistical prediction problems generally require only one representative solution, rather than a list of all solutions.

## 4. Implementation in R and comparison to other packages

The algorithm in this paper has been programmed in R (R Foundation 2019) as the function `HDR.discrete` in the `stat.extend` package (O'Neill and Fultz 2020). This package computes canonical smallest covering regions for a range of probability distributions, including distributions in base R and some extension packages. To accord with common terminology, the package calls the output a HDR although it is actually a CSCR, and only the weak highest density condition is required for the region; hereafter, we will call it a HDR to accord with this terminology. (Other literature on this topic does not discuss the concept of a CSCR at all, so this is a distinction that is not generally considered in these problems. We use the concept here to obtain the smallest sized set with a stipulated minimum coverage probability.)

The `stat.extend` package includes a range of functions of the form `HDR.xxxx` where the suffix refers either to a density shape (e.g., `monotone`, `unimodal`, `bimodal`, etc.) or a distributional family (e.g., `pois`, `binom`, `hyper`, etc.). The `HDR.discrete` function is a general optimisation function used in cases where we have a discrete distribution with known density `f` but with no known structural properties that would assist the optimisation. The inputs for the function are shown in Table 1 below, and the output of the function is a `'hdr'` object containing the highest density region and some information about the optimisation. (This object type has a special user-friendly print method in the package.) All of the inputs appear as inputs in Algorithms 1-2, except for the `'distribution'` input, which is used to give a description of the HDR in the output of the function.

| Table 1: Inputs for the `HDR.discrete` function | |
|---|---|
| **Input** | **Type** |
| `cover.prob` | The coverage probability — real value between zero and one. |
| `f` | The density function for the distribution. |
| `supp.min = -Inf`<br>`supp.max = +Inf` | The bounds on the support of the distribution — real values. Default values impose no restriction on bounds |
| `E = NULL` | The sequence function — a bijective function from the set of positive integers to the set of all integers (or some subset of the integers that covers the support). Default mapping covers all the support (for integers up to the maximum value representable in R). |
| `distribution` | The name of the distribution — character string. |

In the code below we show an example using this function to compute the HDR for a discrete distribution composed of a mixture of Poisson distributions. (Our code is shown in black and the output is shown in blue.) This distribution is bimodal and has countably infinite support. We program this distribution in standard syntax as the **poismix** distribution and we use the **HDR.discrete** function to compute the HDR. In the code shown we specify the minimum coverage probability with **cover.prob**, the minimum of the support with **supp.min**, and we use the default mapping over the support for the computation. We can also see that the description in the **distribution** input is used in the output to describe the HDR.

```
#Generate a discrete mixture of Poisson distributions
dpoismix <- function(x) { 0.3*dpois(x, lambda = 12) +
                          0.3*dpois(x, lambda = 28) +
                          0.4*dpois(x, lambda = 40) }

#Generate the 90% HDR using stat.extend
library(stat.extend)
HHH <- HDR.discrete(cover.prob = 0.9,
                    f = dpoismix, supp.min = 0,
                    distribution = 'a Poisson mixture distribution')

        Highest Density Region (HDR)

90.63% HDR for a Poisson mixture distribution
Computed using discrete optimisation with minimum
coverage probability = 90.00%

7..17, 21..47

#Reformat the HDR into tabular form
reformat(HHH)

            Lower Upper    LC     RC
Interval[1]     7    17 closed closed
Interval[2]    21    47 closed closed
```

The output from the **HDR.discrete** function gives the HDR in a user-friendly format, but we can also use the **reformat** function to reformat it as a table showing the intervals in the HDR. In this case the HDR is composed of the union of two intervals, due to the bimodality of the distribution. The HDR is illustrated in Figure 3 below and the author has confirmed that it is an accurate computation that meets all the properties of a CSCR.

Page 16 of 30

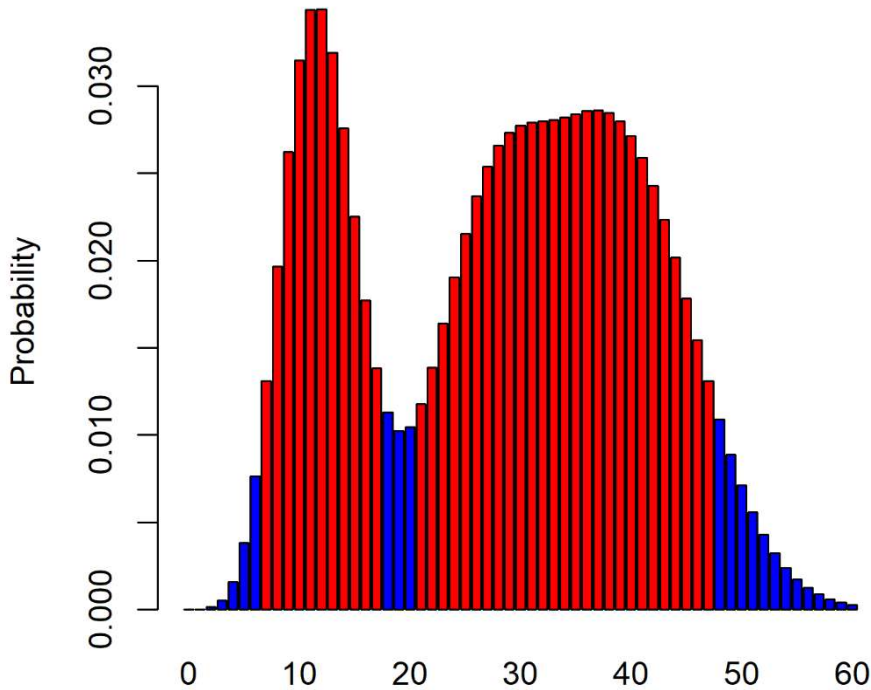

**FIGURE 3:** The 90% HDR for a bimodal Poisson mixture distribution

(Computed using the `HDR.discrete` function in the `stat.extend` package;
$\mathcal{H} = \{7, ..., 17\} \cup \{21, ..., 47\}$ with actual coverage probability $\mathbb{P}(\mathcal{H}) = 0.9063$)

Multimodal distributions and distributions with infinite support both present a challenge for computation methods for HDRs. We can see from the output and the accompanying plot that our method implemented in the `stat.extend` package successfully computes the HDR despite the multimodality and infinite support of the distribution used in our example. Other `R` packages can produce HDRs, but unfortunately none of these is able to correctly compute this region for an arbitrary discrete distribution. At the time of writing, the present author is aware of four other `R` packages that compute HDRs for known distributions: the `hdrcde` package (Hyndman, Einbeck and Wand 2018), the `pdqr` package (Chasnovski 2019), the `HDInterval` package (Meredith and Kruschke 2019), and the `mclust` package (Fraley *et al*. 2020). The package and function descriptions in the documentation for these packages do not explicitly restrict their scope to continuous distributions, but the coding suggests that they were designed with continuous distributions in mind. In Table 2 we show the descriptions of each of the HDR functions we will examine in our comparative analysis; we note that none of the descriptions restrict the scope to continuous distributions. These packages can be applied to discrete distributions (in some cases using simulation or manual manipulations) but it is unclear to the present author whether they were designed with the latter purpose in mind. In any case, since these are presently the only other **R** packages containing functions for HDRs, it is useful to see whether or not they can handle discrete distributions adequately. This will provide users with information on performance for discrete distributions.

| R Package | Function | Description |
|---|---|---|
| `stat.extend` | `HDR.discrete` | This function computes the highest density region (HDR) with support on the integers. The distribution can be any discrete distribution concentrated on the integers — it does not have to have any shape properties for the function to work. The user must give the density function 'f' for the distribution.<br><br>To improve the search properties of the algorithm, the user can also give lower and upper bounds for the support of the distribution if these are available.<br><br>(Warning: If the user specifies incorrect bounds on the support, that do not contain the full support of the distribution, then the algorithm may continue to search without end, in which case the function will not terminate. Similarly, if the user specifies a sequence function E that is not a proper bijection to the integers then the algorithm may continue to search without end, in which case the function will not terminate.)<br><br>The output of the function is a 'hdr' object containing the HDR for the discrete distribution. |
| `HDInterval` | `hdi` | Calculate the highest density interval (HDI) for a probability distribution for a given probability mass. This is often applied to a Bayesian posterior distribution and is then termed "highest posterior density interval", but can be applied to any distribution, including priors.<br><br>The function is an S3 generic, with methods for a range of input objects. |
| `pdqr` | `summ_hdr` | summ_hdr() computes a Highest Density Region (HDR) of some pdqr-function for a supplied level: a union of (closed) intervals total probability of which is not less than level and probability/density at any point inside it is bigger than some threshold (which should be maximum one with a property of HDR having total probability not less than level). This also represents a set of intervals with the lowest total width among all sets with total probability not less than a level. |
| `hdrcde` | `hdr` | Calculates highest density regions in one dimension |
| `mclust` | `hdrlevels` | Compute the levels of Highest Density Regions (HDRs) for any density and probability levels. |

**TABLE 2:** Descriptions of HDR functions in **R** packages (taken from https://rdrr.io/)

In Table 3 we illustrate the functions in these packages using the Poisson mixture example (full code available in the supplementary materials). We compute the HDR using each available function/method; further details are given below. In cases where simulation is needed, we have had to make an applied judgment about the number of simulations to use, and so higher/lower accuracy could be obtained by using more/less simulations than we have used.[6] In cases where the output is a continuous HDR we round to discrete boundaries.

| R Package | Function and method | Computed region | Accurate? |
|---|---|---|---|
| `stat.extend` | `HDR.discrete` | 7 … 17<br>21 … 47 | Yes |
| `HDInterval` | `hdi`<br>`(using quantile function)` | 8 … 46 | No |
| `HDInterval` | `hdi`<br>`(using simulated data*)` | 6.82334 … 17.88118<br>21.12481 … 47.51618 | No** |
| `HDInterval` | `hdi`<br>`(manual computation using`<br>`estimated density cut-off)` | 7 … 17<br>22 … 47 | No |
| `pdqr` | `summ_hdr` | 7 … 17<br>21 … 47 | Yes |
| `hdrcde` | `hdr`<br>`(using density list)` | 6.75725 … 17.81372<br>21.00000 … 47.59820 | Maybe** |
| `hdrcde` | `hdr`<br>`(using simulated data*)` | 6.72782 … 17.86172<br>21.00000 … 47.50582 | Maybe** |
| `hdrcde` | `hdr`<br>`(manual computation using`<br>`estimated density cut-off)` | 7 … 17<br>21 … 47 | Yes |
| `mclust` | `hdrlevels`<br>`(manual computation using`<br>`estimated density cut-off)` | 7 … 17<br>21 … 47 | Yes |

\* This calculation used $10^6$ simulations from the true probability distribution.

\*\* In these cases the region is estimated from a continuous density, so it is sensible to convert to give integer boundaries. There are two plausible methods of rounding that one could employ here: either round all boundaries to the nearest interval; or round all boundaries "inward" to include only the integers that are inside the real intervals.

**TABLE 3:** Computed 90% HDRs for the Poisson mixture distribution (shown in Figure 3)

---

[6] Both the **`HDInterval`** and **`hdrcde`** packages could compute the HDR more accurately if we use the simulation method and use a larger number of simulations. We think we have used a reasonable number of simulations in our analysis, so the inaccuracy of the result still constitutes a drawback of the methods in those packages.

As can be seen from Table 3, none of the other packages (other than **stat.extend**) correctly compute HDRs for an arbitrary discrete distribution, but the **pdqr** package comes close. Some caveats on the comparisons in our table are appropriate. In fairness to the other packages, it is not clear whether they were designed to deal with discrete distributions; most of the functions appear to have been designed to compute HDRs for *continuous* distributions, so here we are shoe-horning some of these functions into an alternative. Moreover, the simulation method in the **HDInterval** and **hdrcde** packages could compute the desired HDR more accurately if we were to use a larger number of simulations, so our results here are heavily dependent on our own choice of the number of simulations. We think we have used a reasonable number of simulations in our analysis, so the inaccuracy of the results shown in Table 3 still constitutes a drawback of the methods in those other packages. Moreover, as will be seen in the descriptions below, it is possible to "break" these algorithms by using more difficult discrete distribution.

**Computation in the HDInterval package:** The **hdi** function in this package allows a range of inputs, including the quantile function of the distribution or a simulated density function. Unfortunately, the function can only accommodate "split" regions when the input is a density function taken from simulated data.[7] Consequently, one can either obtain an optimisation that uses the true quantile function but does not allow split intervals (i.e., optimises only over single connected intervals), or one can obtain an approximate HDR that allows split intervals using the density of simulated values. The former method uses discrete calculus to optimise a single connected interval, while the latter method employs a grid approximation which estimates the vertical density cut-off for the HDR and applies this to the grid of values. Alternatively, one can use the **hdi** function to estimate the density cut-off and then manually compute the values that have a density at least as large as this mass (though this latter method is not automated).

**Computation in the pdqr package:** The **summ_hdr** function in this package allows the user to compute the HDR for an input distribution in the custom format required by the package. The distribution format in the package records discrete distributions as objects where the mass function is recorded over a finite range (rather than as a function), with this finite range set to encapsulate most of the probability mass of the distribution. In the present case the HDR falls within the recorded range, so the function correctly computes the HDR. However, it is possible to "break" the limitations of the distribution object by setting a high coverage probability that requires the search to go into the tails of the distribution. Going near to the extreme end of the range gives inaccurate results for the HDR and going beyond the range limits of the distribution object gives highly inaccurate results. In the special case where the coverage probability is set equal to one the package gives highly inaccurate results for the HDR computation.

---

[7] The documentation for the package warns the user of this restriction. In the case where the input is a simulated density the user can set **allowSplit = TRUE** to allow splits in the computed region.

**Computation in the `hdrcde` package:** The `hdr` function in this package uses a grid method which estimates the vertical density cut-off for the HDR and applies this to a grid of values to give an approximate HDR. The function can either take a set of data or an input density set as a list showing a finite number of values and their density. For a discrete distribution this means that you can give the density over most but not all of the distribution. In either case the function gives an approximate HDR based on the grid method. Alternatively, one can use the `hdr` function to estimate the density cut-off and then manually compute the values that have a density at least as large as this mass (though this latter method is not automated). This package also exhibits some undesirable behaviour when the coverage probability is increased to high levels that go far into the tails of a discrete distribution. In these extremes, increasing the coverage probability can actually reduce the computed region, owing to idiosyncrasies in the grid approximation to the true distribution.

**Computation via estimated density cut-offs in various packages:** Several of the packages examined here compute estimates of the density cut-off for the HDR, and this can be used to manually compute a discrete HDR over a finite section of the discrete distribution. This is done in the `HDInterval` package, the `hdrcde` package and the `mclust` package. (Note that the last of these packages does not compute HDRs; it only computes the estimated density cut-off for a HDR.) This method actually performs quite well in this example; the `hdrcde` and `mclust` packages lead to an estimated density cut-off that is within the correct range to give an accurate HDR for the discrete distribution (in our computations both of these packages gave the density cut-off equal to the smallest density of the true HDR). This allows the user to compute the correct HDR using manual derivation from the estimated density cut-off. One drawback here is that the method is not automated, and it requires the user to manually compare the mass at a pre-specified set of values to the estimated cut-off value.

This particular method works on the current example, but unfortunately it does not always work in these packages. One way it can fail is when we have a distribution that has a spike in its density that is far out in the tails, and this spike is not detected within the finite range of values examined by the algorithms used in the above packages. In the method we have expounded in this paper, such cases are handled by ensuring that our search set meets the outer-boundedness condition, which ensure that there cannot be a density spike in the tails of sufficiently large magnitude to affect the HDR. The methods used in the above packages search over a wide range, but they do not always continue their search until this condition is met. Consequently, it is possible to "break" any of the packages by using a pathological discrete distribution that has a hidden spike so far out in the tails that it is out of the search range. The `HDR.discrete` function in the `stat.extend` package does not suffer from this deficiency, and will always continue searching until the true HDR is found. (Its only search limitation in this respect is imposed by the numerical limits of `R`.)

Other packages for computing HDRs give approximate outputs for discrete distributions, but these do not replicate the exact result that is computed by the method expounded in this paper. Unlike the `stat.extend` package, the other packages examined here accommodate highly general distribution inputs, so it is perhaps not surprising that they do not exactly compute HDRs for discrete distributions. We hasten to stress that these other `R` packages are highly sophisticated in their ability to accommodate a wide range of distributions, and so the lack of accuracy in the present case is a trade-off for a high level of general applicability to a wide range of distribution inputs. Nevertheless, the present analysis shows that there is a useful "gap" in existing statistical software that can be filled by the computation method we describe in this paper. Our method rapidly computes the exact HDR (what we have described as a canonical smallest covering region in the initial sections) and it does this for any discrete distribution, including in cases with multimodality and infinite support.

**6. Conclusion and future work**

We hope that the present algorithm and its implementation in the `stat.extend` package will assist users in situations where they wish to compute an exact HDR from a difficult discrete distribution. Computation of HDRs for discrete distributions has been a problem that has not received much attention in the statistical literature, and we hope that this paper assists in giving some underpinning theoretical results and a useful computational method.

The above analysis is for a probability distribution with known mass function concentrated on a known countable set. A natural extension is to consider the case where we are dealing with an unknown distribution concentrated on a known countable set. In this case, an estimated HDR can be obtained by a simple two-part process, first estimating the discrete probability distribution, and then applying our algorithm to the estimated probability mass function to produce the estimated HDR.

When forming an estimated HDR by this process, it is desirable to use an estimator for the mass function that has a reasonable degree of "smoothness" to avoid an excessively "spikey" estimate of the HDR for small samples. In particular, we recommend that estimation with the empirical distribution be avoided, since the resulting HDR is highly volatile and "spikey" for small to moderate sample sizes. Better results are obtained using of discrete kernel estimators (see e.g., Aitken 1983; Rajagopalan and Lall 1995; Kokonendji and Kiessé 2011; Kiessé 2017; Racine, Li and Yan 2020), which smooth the estimated mass function in various ways, yielding non-zero estimated mass values for outcomes that are not in the sample. Using a mass function estimator that has a degree of "smoothing" gives a resulting estimator for the HDR that is less volatile, and more likely to yield an estimate consisting of roughly the same number of intervals as the true HDR. This is particularly so in small to moderate samples where the data may be sparse over the support.

Nonparametric estimation of HDRs for continuous distributions has been studied extensively in both univariate and multivariate settings (see e.g., Hartigan 1987; Polonik 1995; Tsybakov 1997; Rigollet and Vert 2009; Samworth and Wand 2010; Lei, Robins and Wasserman 2013). This has included analysis of convergence rates, bandwidth selection, and optimal choice of plug-in estimators (Rigollet and Vert 2009; Samworth and Wand 2010; Doss and Weng 2018; Baíllo, Cuesta-Albertos and Cuevas 2001). Detailed examination of appropriate estimators for this task in a discrete setting is beyond the scope of the present paper and the present incarnation of the `stat.extend` package. As a topic for future work it would be useful to examine the convergence properties of discrete HDR estimators formed by applying the algorithm in the present paper to various discrete kernel estimators for discrete mass functions.

Aside from the tricky case of a HDR with coverage probability one (where there is no algorithm that can guarantee finding the correct support of the distribution if it is not already known), our algorithm is guaranteed to compute the HDR accurately, subject only to limitations in the computational power of the machine. In particular, the outer-boundedness condition in our analysis ensures that we will always be able to detect spikes in the mass function occurring out in the tails of the distribution. This is a valuable theoretical advance over existing HDR algorithms, which can be "broken" by these difficult cases.

**Acknowledgement**

The author would like to thank two anonymous referees at *Computational Statistics* for helpful comments on a previous version of this paper, including some suggestions for expanding the scope of discussion in the paper. The suggestions of these referees improved the present paper and I appreciate their work in conducting the review.

# Appendix: Proof of Theorems

In this appendix we set out proofs of the theorems in the main body of the paper and also some associated lemmata.

**LEMMA 1:** Suppose that $\mathbb{P}$ is a discrete probability measure concentrated on a countable set $\Omega$ and consider a set $\mathcal{H}$ that satisfies the following conditions:

$$\begin{aligned}
&\text{(Minimum coverage)} & &\mathbb{P}(\mathcal{H}) \geq 1 - \alpha, \\
&\text{(Moderation)} & &\mathbb{P}(\mathcal{J}) < 1 - \alpha \text{ for all } \mathcal{J} \subset \mathcal{H}, \\
&\text{(Highest density)} & &\nabla(\mathcal{H}) \geq \Delta(\Omega - \mathcal{H}).
\end{aligned}$$

For any set $\mathcal{A}$ with $|\mathcal{A}| < |\mathcal{H}|$ we have $\mathbb{P}(\mathcal{A}) < 1 - \alpha$.

**PROOF OF LEMMA 1:** Since $|\mathcal{A}| < |\mathcal{H}|$ it is possible to choose a strict subset $\mathcal{H}_* \subset \mathcal{H} - \mathcal{A}$ with $|\mathcal{H}_*| = |\mathcal{A} - \mathcal{H}|$. (Note that if $\mathcal{A} \subset \mathcal{H}$ then $\mathcal{H}_*$ is the empty set, but this does not invalidate our working.) We will show that $\mathbb{P}(\mathcal{A}) < 1 - \alpha$ by showing that if we "swap out" the points in $\mathcal{A} - \mathcal{H}$ with the points in $\mathcal{H}_*$ then this cannot increase the probability of the set. First, using the highest density condition on $\mathcal{H}$, and the fact that $\mathcal{H}_* \subseteq \mathcal{H}$, we have:

$$f(x) \leq f(x') \qquad \text{for all } x \in \mathcal{A} - \mathcal{H} \text{ and } x' \in \mathcal{H}_*.$$

Since $|\mathcal{H}_*| = |\mathcal{A} - \mathcal{H}|$ we then have:

$$\begin{aligned}
\mathbb{P}(\mathcal{A}) = \sum_{x \in \mathcal{A}} f(x) &= \sum_{x \in \mathcal{A} \cap \mathcal{H}} f(x) + \sum_{x \in \mathcal{A} - \mathcal{H}} f(x) \\
&\leq \sum_{x \in \mathcal{A} \cap \mathcal{H}} f(x) + \sum_{x \in \mathcal{H}_*} f(x) \\
&= \sum_{x \in \mathcal{A} \cap \mathcal{H} \cup \mathcal{H}_*} f(x) \\
&= \mathbb{P}(\mathcal{A} \cap \mathcal{H} \cup \mathcal{H}_*).
\end{aligned}$$

Now, noting that $\mathcal{A} \cap \mathcal{H}$ and $\mathcal{H}_*$ are disjoint sets and $|\mathcal{A}| < |\mathcal{H}|$, we also have:

$$|\mathcal{A} \cap \mathcal{H} \cup \mathcal{H}_*| = |\mathcal{A} \cap \mathcal{H}| + |\mathcal{H}_*| = |\mathcal{A} \cap \mathcal{H}| + |\mathcal{A} - \mathcal{H}| = |\mathcal{A}| < |\mathcal{H}|.$$

Since $\mathcal{A} \cap \mathcal{H}$ and $\mathcal{H}_*$ are both subsets of $\mathcal{H}$ this implies that $\mathcal{A} \cap \mathcal{H} \cup \mathcal{H}_* \subset \mathcal{H}$ (i.e., this set is a strict subset of $\mathcal{H}$). The moderation condition gives $\mathbb{P}(\mathcal{A}) \leq \mathbb{P}(\mathcal{A} \cap \mathcal{H} \cup \mathcal{H}_*) < 1 - \alpha$ which establishes the result. ∎

**LEMMA 2:** Suppose that $\mathbb{P}$ is a discrete probability measure concentrated on a countable set $\Omega$ and consider a set $\mathcal{H}$ that satisfies the following condition:

$$\text{(Highest density)} \qquad \nabla(\mathcal{H}) \geq \Delta(\Omega - \mathcal{H}).$$

For any set $\mathcal{A}$ with $|\mathcal{A}| \leq |\mathcal{H}|$ we have $\mathbb{P}(\mathcal{A}) \leq \mathbb{P}(\mathcal{H})$.

**PROOF OF LEMMA 2:** Consider an arbitrary set $\mathcal{A}$ with $|\mathcal{A}| \leq |\mathcal{H}|$. Using the highest density condition on $\mathcal{H}$ we have:

$$f(x) \leq f(x') \qquad \text{for all } x \in \mathcal{A} - \mathcal{H} \text{ and } x' \in \mathcal{H} - \mathcal{A}.$$

Since $|\mathcal{A}| \leq |\mathcal{H}|$ we have $|\mathcal{A} - \mathcal{H}| \leq |\mathcal{H} - \mathcal{A}|$ which then gives:

$$\begin{aligned}
\mathbb{P}(\mathcal{A}) = \sum_{x \in \mathcal{A}} f(x) &= \sum_{x \in \mathcal{A} \cap \mathcal{H}} f(x) + \sum_{x \in \mathcal{A} - \mathcal{H}} f(x) \\
&\leq \sum_{x \in \mathcal{A} \cap \mathcal{H}} f(x) + \sum_{x \in \mathcal{H} - \mathcal{A}} f(x) \\
&= \sum_{x \in \mathcal{H}} f(x) \\
&= \mathbb{P}(\mathcal{H}),
\end{aligned}$$

which was to be shown. ∎

**PROOF OF THEOREM 1:** To establish the equivalence asserted in the theorem we must establish that any set meeting the conditions of the theorem is a canonical smallest covering region ($\Longrightarrow$) and any canonical smallest covering region meets the conditions of the theorem ($\Longleftarrow$).

($\Longrightarrow$) Consider an arbitrary set $\mathcal{H}$ that satisfies the three conditions shown in the theorem. Since the coverage condition is already present in the conditions, we have $\mathcal{H} \in \mathfrak{C}(\alpha)$. Application of Lemma 1 shows that $\mathcal{A} \notin \mathfrak{C}(\alpha)$ for any set $\mathcal{A}$ with $|\mathcal{A}| < |\mathcal{H}|$, which means that:

$$|\mathcal{H}| = \min_{\mathcal{S} \in \mathfrak{C}(\alpha)} |\mathcal{S}| = R(\alpha).$$

This ensures that our region meets the constraint in the optimisation problem defining the canonical smallest covering region. Application of Lemma 2 then shows that $\mathbb{P}(\mathcal{A}) \leq \mathbb{P}(\mathcal{H})$ for any set $\mathcal{A}$ with $|\mathcal{A}| \leq |\mathcal{H}|$ (i.e., any $\mathcal{A} \in \mathfrak{C}(\alpha)$). So $\mathcal{H}$ has maximum probability among sets meeting the optimisation constraint, which shows that it is a solution to the optimisation problem defining a canonical smallest covering region.

($\Longleftarrow$) Consider a canonical smallest covering region $\mathcal{H}$. To simplify the proof, we will first dispose of the two extreme cases. In the case where $\alpha = 0$ the class $\mathfrak{C}(\alpha)$ contains all supersets of the support of the distribution. The support of the distribution is a canonical smallest covering region; if this is countably infinite then any countable superset is also a canonical smallest covering region. Any set of this kind clearly obeys the coverage requirement and the highest density requirement (since all outside element have zero probability), but the latter does not obey the moderation requirement.[8] In the case where $\alpha = 1$ the class $\mathfrak{C}(\alpha)$ contains all subsets of the sample space. The empty set is the only canonical smallest covering region. This set trivially obeys each condition in the theorem (the highest density condition being satisfied by the convention that the infimum over the empty set is infinite).

Now that we have disposed of these extreme cases, we will consider the more common case where $0 < \alpha < 1$. Since $\mathcal{H}$ is assumed to be a canonical smallest covering region, it satisfies the optimisation requirement:

$$\mathbb{P}(\mathcal{H}) = \max_{|\mathcal{S}| \leq R(\alpha)} \mathbb{P}(\mathcal{S}).$$

Since $\alpha > 0$ we know that there exists a finite set $\mathcal{C}$ with $\mathbb{P}(\mathcal{C}) \geq 1 - \alpha$ and so the class $\mathfrak{C}(\alpha)$ has at least one member that is finite. In particular, since $R(\alpha) = \min_{\mathcal{S} \in \mathfrak{C}(\alpha)} |\mathcal{S}|$ there must exist a finite set $\mathcal{D}$ with $|\mathcal{D}| = R(\alpha) < \infty$ and $\mathbb{P}(\mathcal{D}) \geq 1 - \alpha$. Thus, we have:

$$\mathbb{P}(\mathcal{H}) = \max_{|\mathcal{S}| \leq R(\alpha)} \mathbb{P}(\mathcal{S}) \geq \mathbb{P}(\mathcal{D}) \geq 1 - \alpha.$$

This establishes the minimum coverage condition. Now, since $|\mathcal{H}| \leq R(\alpha) < \infty$, it follows that for all strict subsets $\mathcal{J} \subset \mathcal{H}$ we have $|\mathcal{J}| < R(\alpha)$ and so by the definition of $R(\alpha)$ we must have $\mathcal{J} \notin \mathfrak{C}(\alpha)$. This means that $\mathbb{P}(\mathcal{J}) < 1 - \alpha$, which establishes the moderation condition in the theorem. It remains only to establish the highest density condition. To do this, suppose —contrary to the theorem— that $\nabla(\mathcal{H}) < \Delta(\Omega - \mathcal{H})$. This requires that there exist elements $x \in \mathcal{H}$ and $x' \in \Omega - \mathcal{H}$ such that $f(x) < f(x')$. Now, construct the new region:

$$\mathcal{H}' = \mathcal{H} - \{x\} + \{x'\}.$$

This region has the same size as $\mathcal{H}$ but we also have:

$$\mathbb{P}(\mathcal{H}') = \mathbb{P}(\mathcal{H}) - f(x) + f(x') > \mathbb{P}(\mathcal{H}),$$

---

[8] Here is where the minor exception in our theorem occurs. If the support is countably infinite then any countable superset of the support is also "of the same size" and yet it may be a strict superset of the support. This means that the superset is a canonical smallest covering region even though it does not obey the moderation condition. Of course, it is possible to exclude "solutions" of this kind by definition if preferred (i.e., set the support of the distribution as the only canonical smallest covering region in this case).

which means that $\mathcal{H}$ is not a canonical smallest covering region. Since this is in contradiction to our initial conditions, we have established by contradiction that the highest density condition holds. This completes the proof. ∎

**PROOF OF THEOREM 2:** To establish the equivalence asserted in the present theorem, we need to establish that any set $\mathcal{H}$ meeting the highest density condition has a search set $\mathcal{S}$ meeting the inner and outer-boundedness conditions ($\Longrightarrow$) and if there is a search set $\mathcal{S}$ meeting the inner and outer-boundedness conditions then the set $\mathcal{H}$ meets the highest density condition ($\Longleftarrow$). We will establish the additional result for finiteness of the search set when dealing with the first of these implications.

($\Longrightarrow$) Consider a set $\mathcal{H}$ meeting the three conditions shown in Theorem 1. We need to establish that the set meets the inner and outer-boundedness conditions. To do this, we first note that for any set $\mathcal{S} \subseteq \Omega$ we have $\Delta(\Omega - \mathcal{H}) \geq \Delta(\mathcal{S} - \mathcal{H})$ so we can apply the highest density condition to get the inner-boundedness condition:

$$\nabla(\mathcal{H}) \geq \Delta(\Omega - \mathcal{H}) \geq \Delta(\mathcal{S} - \mathcal{H}).$$

Since this holds for all $\mathcal{S} \subseteq \Omega$ we can choose $\mathcal{S} = \Omega$ and this ensures that the outer-boundedness is (trivially) satisfied. Moreover, in the case where $\nabla(\mathcal{H}) > 0$ we have $1 - \nabla(\mathcal{H}) < 1$ so there must exist a finite set $\mathcal{H} \subseteq \mathcal{S} \subseteq \Omega$ with $\mathbb{P}(\mathcal{S}) \geq 1 - \nabla(\mathcal{H})$, which gives the outer-boundedness condition for the special case mentioned at the end of the theorem.

($\Longleftarrow$) Consider a search set $\mathcal{H} \subseteq \mathcal{S} \subseteq \Omega$ satisfying the conditions in the present theorem. From the inner-boundedness condition we have $\nabla(\mathcal{H}) \geq \Delta(\mathcal{S} - \mathcal{H})$. From the outer-boundedness condition we also have $\nabla(\mathcal{H}) \geq 1 - \mathbb{P}(\mathcal{S}) = \mathbb{P}(\Omega - \mathcal{S})$. Since $\Omega - \mathcal{H}$ can be partitioned into the disjoint sets $\Omega - \mathcal{S}$ and $\mathcal{S} - \mathcal{H}$ we then have:

$$\nabla(\mathcal{H}) \geq \max\bigl(\Delta(\Omega - \mathcal{S}), \Delta(\mathcal{S} - \mathcal{H})\bigr) = \Delta\bigl((\Omega - \mathcal{S}) \cup (\mathcal{S} - \mathcal{H})\bigr) = \Delta(\Omega - \mathcal{H}).$$

This establishes the highest density condition, which completes the proof. ∎